\documentclass[10pt,conference]{IEEEtran}
\IEEEoverridecommandlockouts
\usepackage{cite}
\usepackage{amsmath,amssymb,amsfonts}
\usepackage{algorithmic}
\usepackage{graphicx}
\usepackage{textcomp}
\usepackage{xcolor}
\usepackage[hidelinks]{hyperref} 
\def\BibTeX{{\rm B\kern-.05em{\sc i\kern-.025em b}\kern-.08em
    T\kern-.1667em\lower.7ex\hbox{E}\kern-.125emX}}
    
\newcommand{\toolname}{ReDSEa}

\begin{document}

\title{ReDSEa: Automated Acceleration of Triangular 
Solver on Supercloud Heterogeneous Systems\\
}

\author{\IEEEauthorblockN{Georgios Zacharopoulos}
\IEEEauthorblockA{\textit{
Computing Systems Lab
}\\
\textit{Huawei }\\
Zurich Research Center\\
Zürich, Switzerland
}
\and
\IEEEauthorblockN{Ilias Bournias}
\IEEEauthorblockA{\textit{
Computing Systems Lab
}\\
\textit{Huawei }\\
Zurich Research Center\\
Zürich, Switzerland
}
\and
\IEEEauthorblockN{Verner Vla\v{c}i\'{c}}
\IEEEauthorblockA{\textit{
Computing Systems Lab
}\\
\textit{Huawei }\\
Zurich Research Center\\
Zürich, Switzerland
}
\and
\IEEEauthorblockN{Lukas Cavigelli}
\IEEEauthorblockA{\textit{
Computing Systems Lab
}\\
\textit{Huawei }\\
Zurich Research Center\\
Zürich, Switzerland
}
}
\maketitle

\begin{abstract} 
When utilized effectively, Supercloud heterogeneous systems have the potential to significantly enhance performance. Our \toolname\ tool-chain automates the mapping, load balancing, scheduling, parallelism, and overlapping processes for the Triangular System Solver (TS) on a heterogeneous system consisting of a Huawei Kunpeng \cite{kunpeng} ARM multi-core CPU and an Ascend 910 \cite{ascend910} AI HW accelerator. 
We propose an LLVM compiler tool-chain that a) leverages compiler analysis
and b) utilizes novel performance models exploring recursive, iterative, and blocked computation models. Our tool-chain facilitates a speedup of up to 16x compared to an optimized 48-core CPU-only implementation.

\end{abstract}


\section{Introduction}
Heterogeneous computing aims to exploit the strengths of diverse processor types and architectures to achieve superior performance, power efficiency, and cost-effectiveness than a homogeneous, general purpose CPU-based system.
The term Supercloud \cite{mccoll2017superclouds} has been used to describe the next generation of the cloud architectures where demanding processing of AI, big data and HPC applications can be supported.
In a large-scale heterogeneous computing environment, such as a Supercloud, high-speed interconnects 
enable different processors to collaborate efficiently using software to distribute and coordinate workloads at various levels of granularity, ranging from low-level hardware acceleration to high-level task scheduling and load balancing.

However, designing complex Supercloud heterogeneous systems poses significant challenges that are both time-consuming and demanding. Engineers must carefully consider physical resource constraints, communication costs, and other factors when deciding which portions of an application should be accelerated on GPUs or hardware accelerators (including TPUs, DPUs, etc.) and which should run on general purpose CPUs.
Moreover, applications in many relevant domains, e.g., Deep Learning, Extended Reality (XR) or Autonomous Vehicles, offer opportunities for various forms of parallel execution, including 
Instruction Level (ILP), Task Level (TLP) and Pipeline (PP) Parallelism. 

These different forms of parallelism, if harnessed for hardware acceleration, can result in significant speedups. Thus, a Design Space Exploration (DSE) methodology based on performance estimation models can be crucial. It can (a) utilize all the forms of parallelism mentioned above, (b) be driven directly by the application source code or a graph representation of a computation model (e.g., recursive, iterative, and blocked), (c) automatically determine the parts of the application that should be accelerated in hardware, and (d) carry out performance estimation.

\section{Related Work}

Monil \textit{et al.} \cite{Lara2022Laris} introduce LaRIS, a portable framework for LAPACK functionalities on Heterogeneous Systems. LaRIS uses IRIS \cite{Lara2022Laris} to
dynamically select the vendor-library kernel and suitable processor architecture at run-time, making the orchestration simpler when different architectures coexist in a single node. Valero-Lara \cite{LaraTrsm} evaluates the use of OpenMP tasking with target GPU offloading as a potential solution for programming productivity and performance on heterogeneous systems. As a test case, the Triangular Solver (TS) routine is used. In both the aforementioned works, there is not a model to determine the refinement level and the expected performance.

In \cite{Relapack}, the authors introduce ReLAPACK, a methodology that makes use of recursive algorithms in order to implement LAPACK functionalities. They claim that, contrary to blocked algorithms, recursive algorithms do not require significant tuning to define the proper granularity/refinement level. Conversely to our work, they target shared memory architectures and not heterogeneous systems.

Compiler-based methodologies \cite{ZacharopoulosNov19, ZacharopoulosApr19, zacharopoulos23trireme} have been proposed to estimate performance on heterogeneous systems employing a general approach and focusing on specific types of parallelism, without exploring different computation models. Finally, DSE methods \cite{ferretti2022graph, BrumarEarlyDSE, ZacharopoulosJul18} that follow strategies for synthesizing accelerators and optimizing them have been explored.

\section{Compiler Tool-chain}

To address all the challenges mentioned in the previous section, we present 
\textbf{\toolname} (\underline{Re}cursive \underline{D}esign \underline{S}pace \underline{E}xploration \underline{A}utomation):
A tool-chain based on the LLVM \cite{LattnerMar04} Compiler Infrastructure (Version 14.0.0).
 
\subsection{Compiler Analysis}

The first stage of our methodology is analyzing the application that is going to be mapped to a heterogeneous system or cloud architecture. The analysis phase includes an automated process that provides our compiler infrastructure with necessary information so as to guide the subsequent steps (Cost Models and Design Space Exploration) of the automation methodology. It can be viewed as the \textbf{input} to our automated system. 
The analysis is performed by generating the respective LLVM-IR \cite{ZacharopoulosMar17} from the application source code (typically written in C, C++). A number of LLVM analysis passes, developed within the scope of \toolname, analyze the intermediate representation of the applications and estimate the latency due to computation of every potential task, or otherwise described, of every node of the data flow graph. Furthermore, the communication cost is estimated by extracting the amount of data that is read and stored by every task. The data requirements, along with the available bandwidth of a target architecture, allows for an estimation of the latency, due to communication. 

\subsection{Cost Models}

To obtain an early evaluation of the potential performance of every computational node, we need to introduce evaluation/cost models that can perform estimations of both computation and communication latency for the components of a Supercloud heterogeneous system, such as CPUs, GPUs, HW accelerators, etc. 

To estimate the latency of a computational node that is mapped to multiple CPUs, GPUs and/or HW accelerators, we need to estimate 1) the latency of the computation on the CPUs, 2) the critical path of the computation that is offloaded to GPUs and/or HW accelerators, 3) the latency to transfer the data from the host main memory to the HW accelerator (Host-to-Device, H2D) and vice versa (Device-to-Host, D2H), and 4) the synchronization/invocation overhead.
$Latency =CPU Computation + HW Computation+Communication+Synchronization/Invocation$

Let $S = \{ S_1, S_2, \ldots, S_N \}$ be a set of nodes (tasks), with associated HW computation latency ($HWcomp_i$), HW communication latency ($HWcomm_i$) and synchronization/invocation overhead ($OVHD_i$). For every node $i$ the cumulative latency will be $HW_i = HWcomp_i + HWcomm_i + Synch_i\ |\ i=1,\dots, N\ $.

\subsection{Design Space Exploration}
Based on the cost models of the previous subsection, a list of candidates for acceleration is generated. 
The selection (branch-and-bound) algorithm recursively explores the subsets of the list of candidates, in a similar manner to the Bron-Kerbosch algorithm \cite{BronKerbosch73}. The output returned is the set
with the highest speedup (minimum cumulative latency) that stays within the user defined resource budget, which is translated as the amount of resources available for hardware acceleration.

\section{Triangular System Solver}

Solving Triangular Systems is a fundamental problem from the dense linear algebra domain. 
In this example, we solve the linear system $Lx = b$, where $L$ is a dense lower-triangular $n \times n$ matrix and $b$ is a dense vector of length $n$. 

\begin{figure}[t]
\centering
\includegraphics[width=0.7\linewidth]{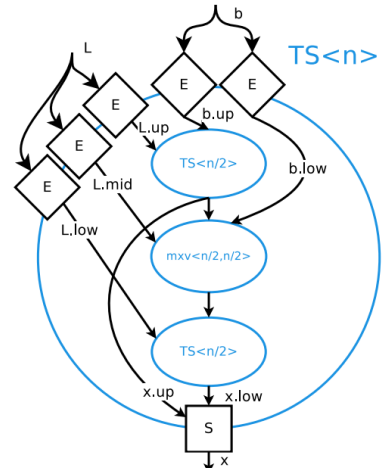}
\vspace{-0.5cm}
\caption{Data Flow Graph (DFG) of the recursive implementation of triangular system solver \texttt{TS<n>} and its refined (decomposed) nodes \texttt{TS<n/2>} and the matrix-vector multiplication update \texttt{mxv<n/2,n/2>}.
}
\vspace{-0.5cm}
\label{fig:ts-rec}
\end{figure}

The solution to this problem, in order to explore multiple levels of granularity, relies on dividing $b$ into an upper
and a lower half and the matrix $L$ into matrices $L.up$, $L.mid$, and $L.low$ corresponding
to the upper left, lower left, and lower right blocks of $L$, as seen in Figure \ref{fig:ts-rec}. $L.up$, $L.low$ are
dense lower-triangular $n/2\times n/2$ matrices and $L.mid$ is a full dense $n/2\times n/2$ matrix. With this analysis, the system can be solved in 3 stages:
\begin{enumerate}
    \item Solve the triangular system $(L.up)(x.up) = b.up$
    \item Update $b.low = b.low - (L.mid)(x.up)$
    \item Solve the triangular system $(L.right)(x.low) = b.low$
\end{enumerate}
The solution to the original system is $x = (x.up, x.low)$.

We extend the problem by solving $n$ linear systems for $n$ different $b$ vectors of size $n$ while keeping the same $L$  lower-triangular ($n \times n$) matrix.

\section{Models of Computation}

Three models of computation are explored within the scope of this work:
Recursive, as seen in \cite{Relapack}, Iterative and Blocked.

\subsection{Recursive}
\label{sec-rec)}

The recursive model of computation offers a decomposition of the initial problem to several tasks, as shown in Figure \ref{fig:ts-rec}. It also provides the opportunity to expose parallelism and finally to explore various levels of granularity in order to make the best architectural decisions and use efficiently any given available software (general purpose CPUs) and hardware accelerators resources. 

In the example of the Triangular System (TS), the initial problem of size $n$ is decomposed to the tasks: \texttt{TS<n/2>}, general matrix-vector multiplication \texttt{gemv<n/2,n/2>} and \texttt{TS<n/2>}, along with the respective dependencies. The gemv task offers data level parallelism and, hence, can be an excellent candidate to be accelerated. As we solve $n$ instances of TS for $n$ vectors of b, the gemv task becomes a matrix-matrix multiplication task and the second stage of the problem is transformed to a general matrix multiplication (gemm) task.

The refined \texttt{TS<n/2>} tasks can be refined as well, so as to expose more parallelism and accelerate a larger part of the initial \texttt{TS<n>}. For every iteration $i$ of the decomposition/refinement of TS, the size of the initial matrix $n \times\ n$ ($i=0$) is decreased to a quarter of the previous iteration (e.g.  $n/2 \times\ n/2$ for $i=1$, $n/4 \times\ n/4$ for $i=2$ etc.).  

The next refinement stage is applied \emph{only} if it benefits performance, i.e., if the following \textbf{condition} is satisfied:
$2 \times TS(i+1) < TS(i)\ |\ i=0,\dots, N\ $\\
We define as $r(i)$ the \textbf{Refinement Level} of iteration $i$.\\
\begin{equation} 
 r(i)=2^i\ |\ i=0,\dots, N\ 
 \end{equation}
The models that estimates the performance, i.e., computation and  communication 
latency, on a heterogeneous architecture are described by the following formulas:
\vspace{-0.5cm}
%
%

\begin{align*}
&Comp(i) = r(i) \times\ TS(i) + \sum_{j=0}^{i-1} r(j) \times\ gemm(j)
\\
&Comm(i) = \sum_{j=0}^{i-1} r(j)\times Comm_{(H{2}D+D2H)}(j)
\end{align*} 

\begin{figure}[t]
\centering
\includegraphics[width=1\linewidth]{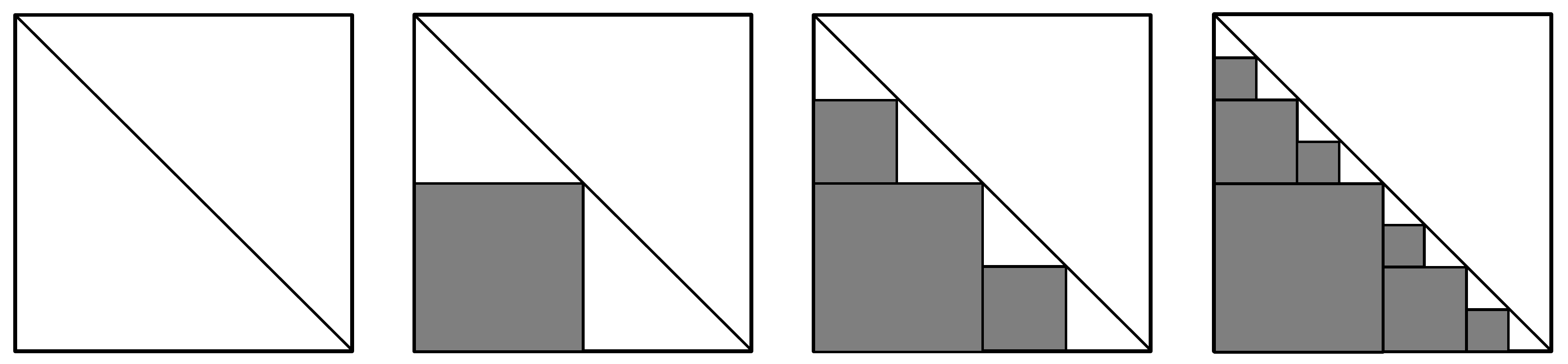}
\caption{Recursive model for Triangular System Solver (Iteration: 0, 1, 2 and 3 Refinement Level: 1, 2, 4 and 8). 
Each grey rectangle (square) represents the gemm computation that is offloaded to an accelerator. The white triangles represent the part of the computation (TS) that remains in the host CPU.}
\label{fig:recursive}
\end{figure}

\subsection{Iterative}
\label{sec-iter)}

Once the granularity of the TS computation that remains on the host has been determined, an iterative model of computation could be offered as an alternative.\\
There are two main reasons to favor an iterative approach over a recursive one: a) It can offer better utilization of the accelerators that are available, as fewer accelerators are dedicated to compute relatively smaller parts of the computation compared to a recursive one which allocates smaller and smaller parts of the computation to be computed by the HW accelerators. 
b) It demands less engineering effort to be implemented while at the same time the performance is not sacrificed, but instead it is kept equal, or slightly better, compared to a recursive one.

\begin{figure}[t]
\centering
\includegraphics[width=1\linewidth]{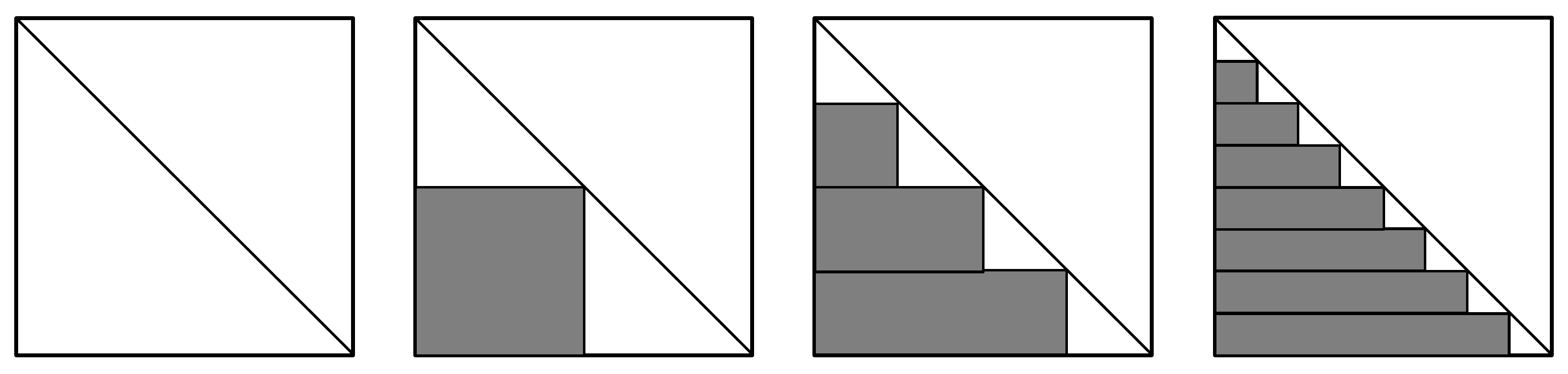}
\vspace{-0.5cm}
\caption{Iterative model for Triangular System Solver (Iteration: 0, 1, 2 and 3 Refinement Level: 1, 2, 4 and 8). 
Each grey rectangle represents the gemm computation that is offloaded to an accelerator. The white triangles represent the part of the computation (TS) that remains in the host CPU.
}
\label{fig:iter}
\end{figure}

\begin{figure}[t]
\centering
\includegraphics[width=1\linewidth]{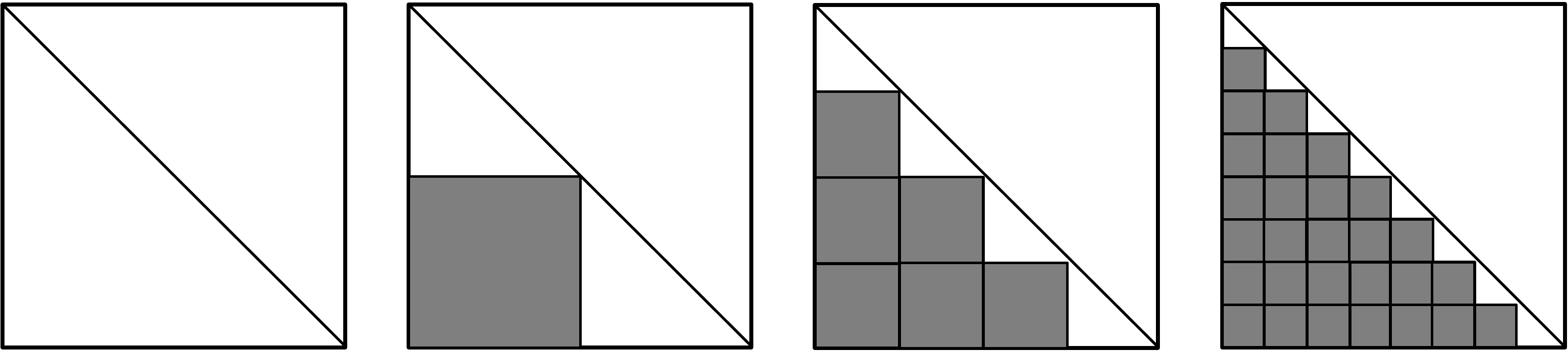}
\caption{Blocked model for Triangular System Solver (Iteration: 0, 1, 2 and 3 Refinement Level: 1, 2, 4 and 8). 
Each grey rectangle (square) represents the gemm computation that is offloaded to an accelerator. The white triangles represent the part of the computation (TS) that remains in the host CPU.
}
\label{fig:blocked}
\end{figure}

The models that estimate the expected performance and communication cost  of the iterative approach are:
\begin{align*}
&Comp(i) = r(i) \times\ TS(i) + \sum_{j=0}^{r(i)-2} gemm(i,j)
\\
&Comm(i) =   \sum_{j=0}^{r(i)-2} (Comm_{H{2}D}(j) + Comm_{D{2}H}(i))
\end{align*}


\subsection{Blocked}

\begin{figure}[b]
\centering
\includegraphics[width=1\linewidth]{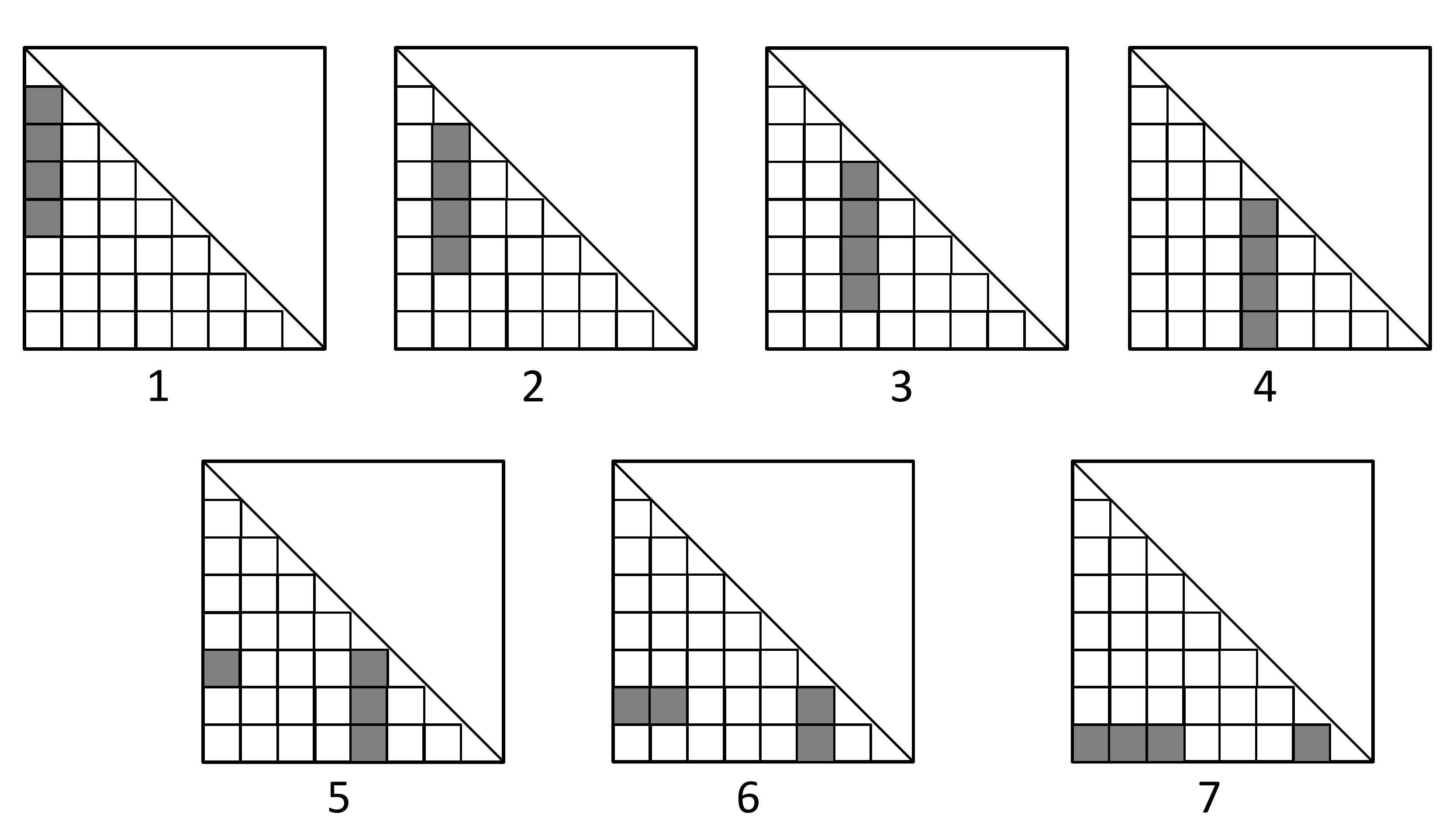}
\caption{Workflow in seven rounds of computation for the blocked model of Triangular Solver (Refinement: $r(3) = 8$, Rounds: $r(3)-1=7$, Blocks: $ \dfrac{r(3)}{2}=4$  ).
}
\label{fig:blocked-flow}
\end{figure}

A blocked approach, as seen in Figure \ref{fig:blocked}, can be used once the level of refinement and granularity of the TS computations that reside at the host CPU has been determined.

The advantages of a blocked model, compared to a recursive or an iterative one, are better overall load balancing, more efficient use of HW accelerators resources, and better scheduling. 

The workflow of the blocked model (Figure \ref{fig:blocked-flow}) allows a computation in rounds offloading equivalent workloads for the acceleration of gemm in every round and using the available resources efficiently.

For every iteration $i$ and a respective refinement level $r(i)$, the number of \textbf{rounds of computation} is $r(i)-1$  and the \textbf{blocks of computation} is $ \dfrac{r(i)}{2}$  per round, 
so at to have equal workloads per round.
Also, the partitioning of the computation into blocks unlocks the potential for parallelism execution with multiple accelerators, requiring less engineering effort compared to the previous models, and it unlocks the option to overlap the acceleration of gemm with the CPU computation of TS.

Thus, the total number of blocks (gemm(i) computations) to be accelerated are
$(r(i)-1) \times \dfrac{r(i)}{2}$ (rounds of computation multiplied by the blocks   of computation per round). In the example of Figure \ref{fig:blocked-flow}, the total number of blocks is  $7 \times\ 4 = 28$.



The  models  that compute  the expected  performance  and communication cost of the blocked approach are:

\begin{align*}
&Comp(i) = r(i) \times\ TS(i) + ((r(i) -1) \times \dfrac{r(i)}{2} ) \times\ gemm(i)
\\
&Comm(i) =   ((r(i) -1) \times \dfrac{r(i)}{2} ) \times\ Comm_{(H2D+D2H)}(i)
\end{align*}

\begin{figure}[t]
\centering
\includegraphics[width=\linewidth]{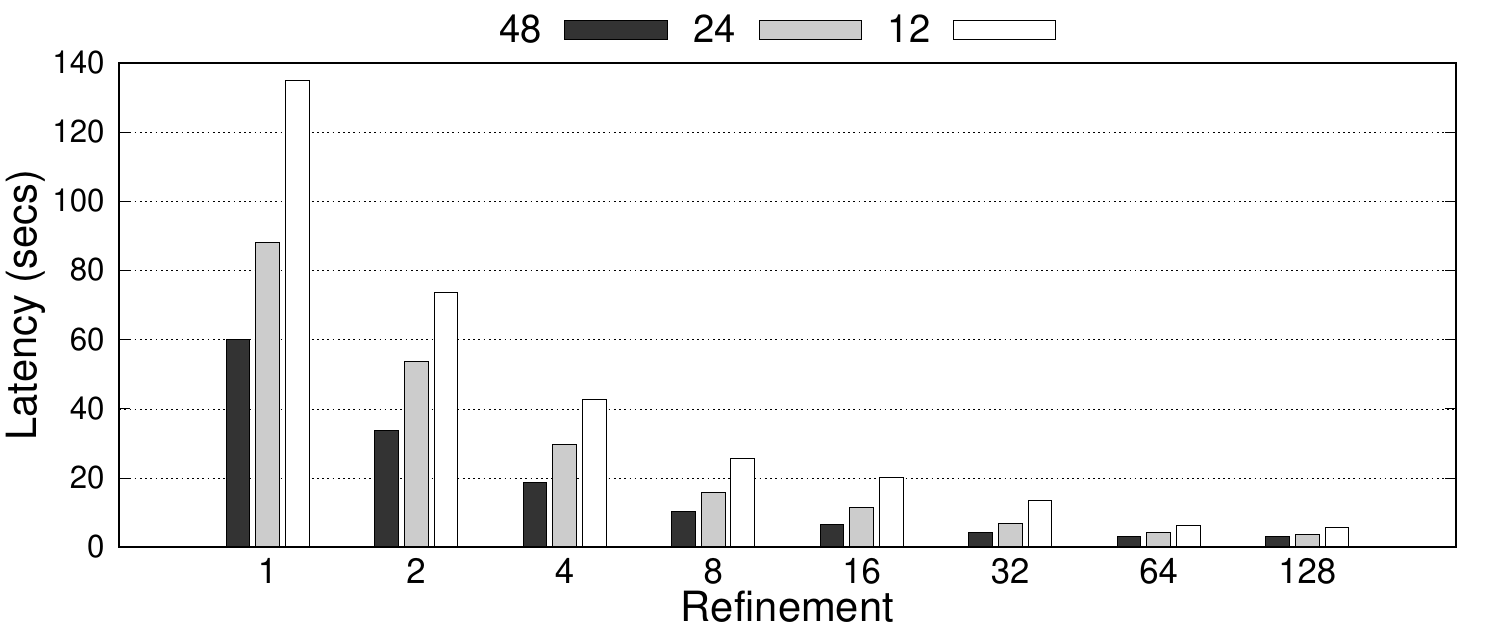}

\includegraphics[width=\linewidth]{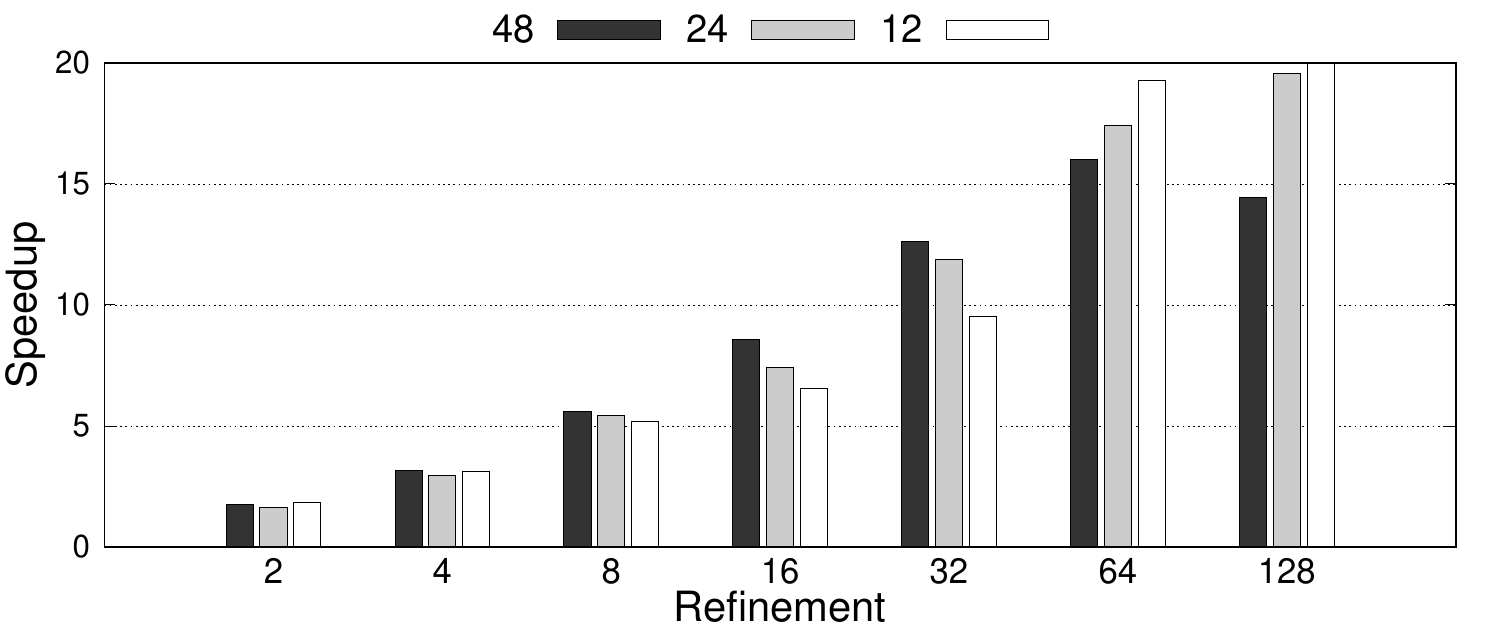}
\vspace{-0.7cm}
\caption {Latency (top) and Speedup (bottom) obtained for Triangular System (TS) Solver increasing the refinement level and gradually offloading more of the initial computation to the Ascend device. The computation remaining in the host CPU was computed using 48, 24 and 12 cores respectively.
}
\vspace{-0.5cm}
\label{fig:ts-speed}
\end{figure}
\section{Experimental Results}
Our experimental setup consists of a general purpose 48-core ARM CPU (Huawei Kunpeng \cite{kunpeng}) and a Huawei Ascend 910 AI processor \cite{ascend910}, that consists of 32 Da Vinci AI cores with a peak performance of 320\,TFLOPS. The two devices communicate via a PCIe bus. 
The results are equivalent for all three computation models explored.
In Figure \ref{fig:ts-speed} (top) the total latency of every heterogeneous design can be seen while the refinement level increases and while using either all available CPU resources (48 cores), half of them (24 cores) or finally 12 of the available CPU cores. By increasing the refinement level, hence offloading a larger part of the computation to Ascend, significant time can be saved even when using fewer CPU cores, e.g., with refinement equal to 32 and 12 CPU cores.

\begin{figure}[t]
\centering
\includegraphics[width=\linewidth]{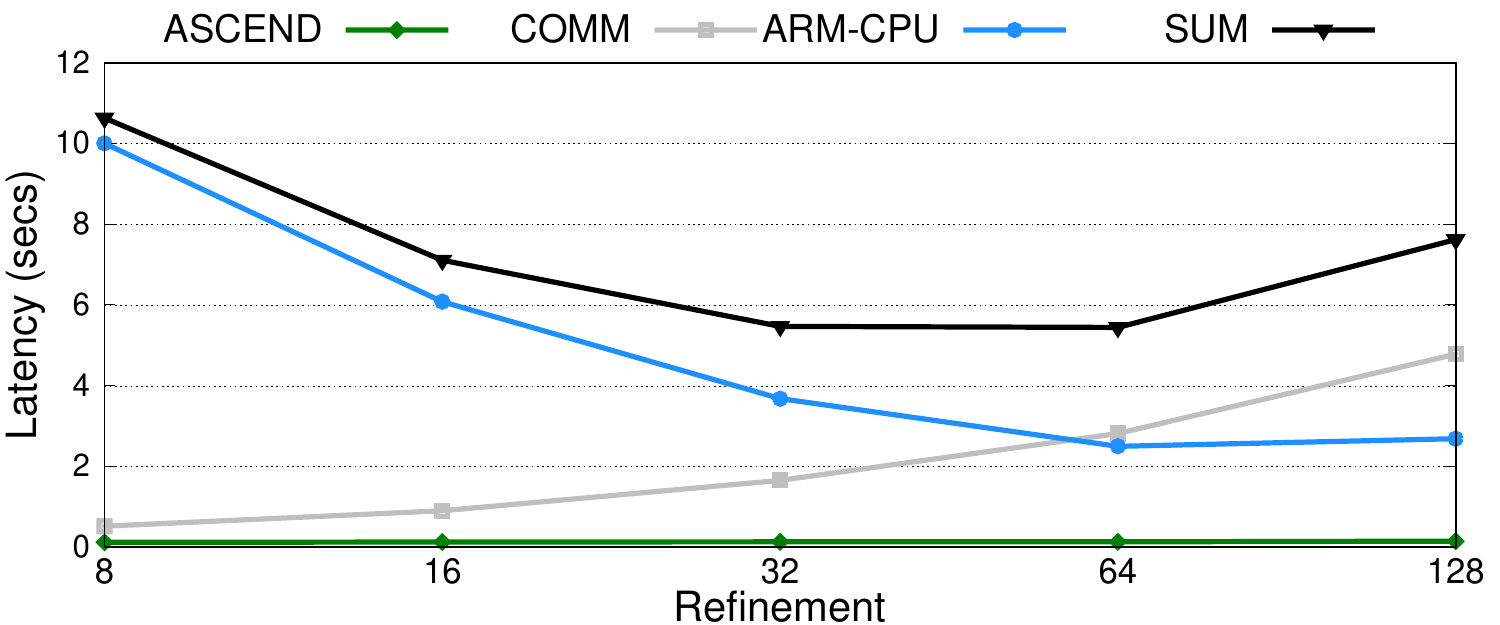}
\vspace{-0.6cm}
\caption {Latency of Ascend accelerator computation, host-to-device and device-to-host communication, ARM CPU (48 cores) computation and their sum.
}
\label{fig:ts-se-hw-comm}
\end{figure}

The respective speedup over an optimized CPU-only baseline, showcased in Figure \ref{fig:ts-speed} (bottom) can be up to a compelling \textbf{16x} using 48 CPU cores (refinement=64). However, the speedup decreases with the next iteration of refinement (128), which also employs finer granularity. This is due to two main factors: a) The CPU-residing part of the TS computation is so fine-grained that it cannot be executed faster than the previous iteration. So, the condition $2 \times TS(i+1) < TS(i)\ |\ i=0,\dots, N\ $, as defined in Section \ref{sec-rec)} is not satisfied and the refining process ends. This can be observed in Figure \ref{fig:ts-se-hw-comm}, where the CPU latency (48 cores) of the last refinement iteration (128) is larger than the previous iteration. b) The communication cost raises substantially while refinement increases. The communication latency between the CPU host and the Ascend device at the last two refinement iterations (64 and 128) surpasses the cost of the CPU computation resulting in significant overhead and halts the potential for more speedup.


\section{Conclusions}


Using the \toolname\ tool-chain, which incorporates performance models derived from recursive, iterative, and blocked computation models, we have managed to automatically map the Triangular Solver (TS) onto a heterogeneous system, consisting of a Kunpeng 48-core ARM CPU and an Ascend AI accelerator device, achieving up to a 16x speedup. We have explored three models of computation with an emphasis on the blocked approach to minimize HW acceleration times compared to the recursive and iterative versions.

\section{Future Directions}

Our objective is to investigate the potential of parallelism and overlapping in the blocked model, aiming to showcase their impact on the overall latency and speedup achieved by the respective designs.
We also plan to 
extend this methodology to other applications in the HPC and AI domains, such as Dense Cholesky Factorization, QR Matrix Factorization, and others. Developing new models or extending existing ones will be necessary to estimate the performance of more applications. Finally, we aim to apply this methodology to more complex heterogeneous and Supercloud architectures.







\bibliographystyle{plain}
\bibliography{conference}


\end{document}